\documentstyle[psfig,graphicx]{mn}

\def\gapprox{\lower.4ex\hbox{$\;\buildrel >\over{\scriptstyle\sim}\;$}}
\def\lapprox{\lower.4ex\hbox{$\;\buildrel <\over{\scriptstyle\sim}\;$}}

\title{The effect of differential refraction on wave propagation in rotating pulsar magnetospheres}

\author[Fussell and Luo]
      {D. Fussell and Q. Luo\\
        School of Physics, The University of Sydney, NSW 2006, Australia\\
}

\date{
          --- Received
         in original form October, 2003
        }

\pubyear{2003}

\begin{document}

\maketitle

\begin{abstract}
Refraction of wave propagation in a corotating pulsar magnetospheric plasma is 
considered as a possible interpretation for observed asymmetric pulse profiles with multiple
components. The pulsar radio emission produced inside the magnetosphere propagates outward 
through the rotating magnetosphere, subject to refraction by the intervening plasma that 
is spatially inhomogeneous. Both effects of a relativistic distribution of the plasma and 
rotation on wave propagation are considered.  It is shown that refraction coupled with rotation 
can produce asymmetric conal structures of the profile. The differential refraction due to 
the rotation can cause the conal structures to skew toward the rotation direction and lead 
to asymmetry in relative intensities between the leading and trailing components. Both of 
these features are potentially observable.
\end{abstract}


\section{Introduction}

Pulsar radio emission is thought to originate in the region deep inside the
pulsar magnetosphere populated with a relativistic electron-positron pair
plasma (e.g. Blaskiewicz, Cordes, \& Wasserman 1991; Melrose 2000). The radio 
waves propagate through the magnetosphere, subject to 
reflection and refraction (e.g. Barnard \& Arons 1986; Petrova 2000;
Fussell, Luo \& Melrose 2003, and hereafter FLM) or absorption (Blandford \& Scharlemann 1976; 
Lyubarskii \& Petrova 1998; Luo \& Melrose 2001; FLM). Since these propagation effects 
can give rise to observable features in the pulse profile and polarization, study 
of them can provide insight to the radiation processes inside the magnetosphere.

In the polar cap model,  relativistic electron positron plasmas are
produced in the cascade above the polar cap (PC) as a result of 
rotation-induced particle acceleration (Sturrock 1971; Ruderman \& Sutherland 1975;
Arons \& Scharlemann 1979; Daugherty \& Harding 1982;  Zhang \& Harding 2000; Hibschman \& 
Arons 2001). Primary electrons (positrons) are accelerated  to ultra-high energies by a 
rotation-induced parallel electric field, emitting high energy photons through curvature 
radiation or inverse Compton scattering. High energy photons decay into electron/positron 
pairs in the strong pulsar magnetic field, forming an outflowing relativistic pair plasma, 
referred to as the pulsar plasma, which has a very broad distribution in parallel momenta.
Although the mechanism for the radio emission is not well understood, it is generally 
believed that the emission is produced from collective plasma processes in the pulsar plasma, 
through either maser emission or plasma instabilities (e.g. Melrose 2000). Regardless of
the specific emission mechanism, the radiation must be produced directly on or converted 
to modes that can escape the magnetosphere to reach the observer.  In the pulsar plasma, there 
are two escape modes, the X mode with polarization perpendicular to the local plane of the 
magnetic field and wave vector, and the LO mode with polarization in this plane (e.g. Kennett, 
Melrose \& Luo 2000, hereafter KML, and references therein). Observations of jumps
between two orthogonal polarizations in the position angle of the radiation
support the hypothesis that the radio emission propagates through the magnetosphere in 
these two orthogonal modes (Stinebring et al. 1984;
McKinnon \& Stinebring 2000). Therefore we only consider propagation of these two modes.

A pulsar plasma is spatially inhomogeneous due to the radial and transverse
dependence of the density in the open field line region. Wave propagation in such an
inhomogeneous medium can be affected by refraction and reflection, which ultimately determines
the emergent pulse profile. Because the dipole field, $B$, decreases in the radial direction
as $B\propto1/R^3$, where $R$ is the radial distance to the star's center, the density of the 
plasma that streams along the field lines must have the same radial dependence, 
$N\propto 1/R^3$, corresponding to the longitudinal characteristic length scale for the density
variation, $L\sim R$, considerably larger than the star's radius. Variation in the plasma 
density in the direction perpendicular to the field lines is mainly due to the electron/positron 
pair cascade being nonuniform across the PC, which gives rise to a much smaller (than $R$) transverse length 
scale of order the radius of the open field line cone. Although the inhomogeneities can be on a still much 
smaller scale, possibly as a result of nonstationary acceleration or highly localized pair 
cascades, we only discuss the case that both radial and transverse scales are much larger 
than the relevant wavelength and that the refraction is the dominant effect that changes 
the ray path. Wave propagation can be then treated in the geometric optics approximation in 
which two orthogonal modes propagate independently of each other (Barnard \& Arons 1986). 

The effect of refraction on wave propagation in the pulsar magnetosphere was discussed in the 
cold plasma approximation by several authors (e.g. Barnard \& Arons 1986; Petrova 2000; 
Weltevrede et al. 2003). Numerical study of electron/positron pair cascades 
above the PC shows that the distribution of a pulsar plasma has a relativistic spread in the 
plasma rest frame (e.g. Daugherty \& Harding 1982; Zhang \& Harding 2000; Hibschman \& Arons 
2001; Arendt \& Eilek 2002). The relativistic distribution strongly modifies the wave properties 
(KML) and therefore a self-consistent model for wave propagation 
needs to include the relativistic effect of the plasma. Past work on refraction of wave propagation
generally ignores rotation and the ray path initially in the field line direction would
remain two dimensional, confined to the plane of the magnetic field lines. The resultant
pulse profile has the same symmetry as the intensity distribution at the emission origin (e.g. 
Petrova 2000). Observations often show pulse profiles with asymmetric multiple components, i.e.
there is asymmetry in intensity as well as location in pulse longitude of conal components
(Lyne \& Manchester 1988; Gangadhara \& Gupta 1998; Gupta \& Gangadhara 2003). It is
believed that distortion of the profile is due to aberration (Blaskiewicz, Cordes \& Wasserman 
1991; Hirano \& Gwinn 2001; Gupta \& Gangadhara 2003) or absorption (Luo \& Melrose 2001; FLM). 
Wave propagation in rotating magnetospheres was recently discussed in FLM with emphasis on 
the cyclotron absorption. Cyclotron absorption including rotation can lead to differential 
absorption producing asymmetric pulse profiles. In their discussion, FLM assumed the resonance 
region to be at a substantial fraction of the light cylinder (the radius at which the
corotation speed equals the light speed, $c$) where refraction can be ignored. 

The purpose of this paper is to explore the effect of refraction on wave propagation in the
relativistic pulsar plasma that corotates with the star and the implication of such effect
for interpretation of pulse profiles. We consider effects of both a 
relativistic distribution of the plasma and rotation. Due to rotation rays that originate from
the leading and trailing components are subject to different refraction. It is suggested that such
differential refraction can significantly distort the pulse profile. Strong refraction occurs in 
the region with a large density gradient, which is referred to as the refraction region and
is located well below the cyclotron resonance region (FLM). To concentrate on the effect of refraction,
one assumes the strong magnetic field approximation, in which the X mode is not affected by the 
medium and propagates approximately in a straight line through the magnetosphere. The LO mode is strongly 
refracted  and is discussed here. Following a similar procedure to that described in FLM, the 
ray path in the rotating medium is evaluated numerically in the geometric optics formalism. The 
emergent pulse profile is obtained at radial distances beyond which the refraction is no longer
effective. 

In Sec. 2, the wave dispersion of relativistic pulsar plasma is summarized with
emphasis on the LO mode in the strong magnetic field approximation; The formalism of ray tracing in 
the rotating magnetosphere is described in Sec. 3. The ray path is obtained by numerically solving
the ray equations including the relativistic distribution and rotation. The result is applied to
interpretation of asymmetry of pulse profiles (Sec. 4). 

\section{Relativistic pulsar plasma}

The wave properties can be obtained based on the relativistic model of a pulsar plasma (e.g.
KML). The distribution of the secondary pairs is broad, characterised by a 
cutoff at the lower energy near the (pair production) threshold and a rapid increase to a 
peak with the Lorentz factor at around $\gamma\sim 10^2$, followed by a decay and then a cutoff 
at the pair producing photon energy of about $\gamma\sim 10^3-10^4$ (Zhang \& Harding 2000; 
Hibschman \& Arons 2001; Arendt \& Eilek 2002).  It is convenient to use the plasma rest frame, i.e.
the center-of-momentum frame with the transform velocity defined by 
$v_s=\langle\gamma v\rangle/\langle\gamma\rangle$, where $\gamma=1/(1-v^2)^{1/2}$, $v$ is the velocity
(in $c$), and $\langle...\rangle$ is the average over the plasma distribution. The distribution in 
the plasma rest frame has a much simpler form, which has an approximately symmetric peak with a 
relativistic spread and can be approximated by the J\"uttner distribution (i.e. a thermal distribution
with a relativistic temperature). In practical calculations of the plasma dispersion it is reasonable
to use a much simpler distribution such as the bell-type distribution (Melrose \& Gedalin 1999).
This latter distribution is adopted here.

\subsection{Relativistic dielectric tensor}

The response of a relativistic plasma is fully determined by the three relativistic 
plasma dispersion functions (RPDFs) $W(z)=\langle \gamma^{-3}(v-z)^{-2}\rangle$,
$R(z)=\langle\gamma^{-1}(v-z)^{-1}\rangle$, and $S(z)=\langle\gamma^{-2}(v-z)^{-1}\rangle$, 
where $z=\omega/k_\Vert c$ is the parallel (to the magnetic field, $\bf B$) phase velocity (in units of $c$), 
$k_\Vert$ is the projection of the wave vector along the magnetic field.

As we are interested in the inner magnetosphere region where refraction is important, we adopt
the low-frequency limit, $\omega\ll\Omega_e$, where $\Omega_e$ is the cyclotron frequency,
and the non-gyrotropic approximation. Only $W(z)$ is required in determining the dispersion. 
Specifically, in the strong magnetic field approximation the dielectric tensor is reduced
to (Melrose \& Gedalin 1999; KML)
\begin{equation}
K_{11}\approx K_{22}\approx1,\ \ 
K_{33}\approx1-{\omega^2_p\over\omega^2}z^2W(z),
\end{equation}
where the off-diagonal components are ignored, the 3-axis is along $\bf B$, the wave vector is in the
1-3 plane at an angle $\theta$ to $\bf B$, $\omega_p=(e^2N/\varepsilon_0m_e)^{1/2}$ is 
the plasma frequency, and $N$ is the total number density of electrons and positrons. 

The properties of $W(z)$ are discussed in detail in KML. 
One of its main features is that the function has a peak near $z=\omega/k_\Vert c\approx 1$, 
which is characterized by the intrinsic relativistic spread $\langle\gamma'\rangle\ge1$
(where $\gamma'=1/(1-{v'}^2)^{1/2}$ is the Lorentz factor in the plasma rest frame) and is not particularly 
sensitive to the specific form of distribution. We consider a 
normalized soft-bell distribution in the plasma rest frame given by
\begin{eqnarray}
F'(u')&=&{15\over16\gamma'_m{v'}^5_m}{(v'_m-v')^2(v'_m+v')^2\over(1-{v'}^2)^2}
\nonumber\\
&&\times H(v'_m-v')H(v'+v'_m),
\label{eq:distribution}
\end{eqnarray}
where $H(x)=1$ for $x>0$ and $H(x)=0$ for $x<0$,
$u'=\gamma' v'$ and $-u'_m\leq u'\leq u'_m=\gamma'_mv'_m$, $\gamma'_m=1/(1-{v'}^2_m)^{1/2}$, 
$v'_m$ is the maximum velocity in the plasma rest frame. Using (\ref{eq:distribution}) $W(z)$
can be derived analytically. Since $F'(u')$ is a Lorentz invariant, substituting $v'=(v-v_s)/(1-v_sv)$ for 
(\ref{eq:distribution}), the corresponding un-normalized distribution in the
pulsar frame, $F(u)=F'(u')$, can be obtained as
\begin{eqnarray}
F(u)&=& {15\Gamma^4_s\over16\gamma_mv^5_m}
{(v_{max}-v)^2(v_{min}+v)^2\over(1-v^2)^2}\nonumber\\
&&\times H(v_{max}-v)H(v-v_{min}),
\end{eqnarray}
where $v_{max}=(v_s+v'_m)/(1+v'_mv_s)$, $v_{min}=(v_s-v'_m)/(1-v'_mv_s)$,
$\Gamma_s=1/(1-v^2_s)^{1/2}$ is the bulk Lorentz factor.

\subsection{The LO mode}

In the strong magnetic field, the dispersion of the X mode is close to that
of vacuum waves and since this mode is not refracted, it is not considered here. 
The dispersion of the LO mode can be written as (Melrose \& Gedalin 1999; 
Melrose, Gedalin, Kennett, Fletcher 1999)
\begin{equation}
{\omega^2\over\omega^2_p}\approx z^2W(z){z^2-1\over z^2-1-\tan^2\theta_{kb}},
\end{equation}
where $\theta_{kb}$ is the propagation angle to the magnetic field, $\bf B$, and
the RPDF in the pulsar frame is given by
\begin{eqnarray}
&&W(z)={15\Gamma^3_s\over16v^5_m\gamma_m(1-z^2)^3}\Biggl\{v_m\Gamma^{-2}_s(z^2-1)
\bigl[8v_sz\gamma^{-2}_m \nonumber\\
&&-(1+v^2_sz^2)(1-3v^2_m)-(3-v^2_m)(z^2+v^2_s)\bigr]\nonumber\\
&&-{\textstyle{1\over4}}\gamma^{-2}_m(1+v_s)^3(z-1)^3\bigl[(1+3v^2_m)(1-v_sz)\nonumber\\
&&-(v_s-z)(3+v^2_m)\bigr]\ln\left|{(1-v_m)(1+v_mv_s)\over(1+v_m)(1-v_mv_s)}\right|\nonumber\\
&&-{\textstyle{1\over4}}\gamma^{-2}_m(1-v_s)^3(z+1)^3\bigl[(1+3v^2_m)(1-v_sz)\nonumber\\
&&+(v_s-z)(3+v^2_m)\bigr]\ln\left|{(1+v_m)(1+v_mv_s)\over(1-v_m)(1-v_mv_s)}\right|\nonumber\\
&&-4\gamma^{-2}_m(v_s-z)(1-v_sz)\left[v^2_m(1-v_sz)^2-(v_s-z)^2\right]\nonumber\\
&&\times\ln\left|1-{2v_m(1-v^2_s)\over
(1-v_mv_s)[(v_m+v_s-z(1+v_mv_s)]}\right|\Biggr\},
\end{eqnarray}
where $v_m$, replacing $v'_m$, is the maximum velocity in the plasma rest frame.
The average Lorentz factor is given by
\begin{eqnarray}
\langle\gamma\rangle&=&
{15\Gamma_s\over16\gamma_mv^5_m}
\Biggl[
-{v_m\gamma^2_m\over 24}(3+4v^2_m-15v^4_m-15v^2_s\nonumber\\
&&+4v^2_mv^2_s+3v^4_mv^2_s)-{\textstyle{1\over 16}}(1+2v^2_m+5v^4_m-5v^2_s\nonumber\\
&&-2v^2_mv^2_s-v^4_mv^2_s)
\ln\left({1-v_m\over 1+v_m}\right)\Biggr].
\label{eq:avgamma}
\end{eqnarray}
In the relativistic limit ($\Gamma_s\gg1$, $\gamma_m\gg1$) the average
Lorentz factor in the pulsar frame can be approximated by
$\langle\gamma\rangle\approx2\Gamma_s\langle\gamma'\rangle$, where
$\langle\gamma'\rangle\approx 5\gamma_m/16$ is the average Lorentz factor in the plasma rest frame.

\begin{figure}
\psfig{file=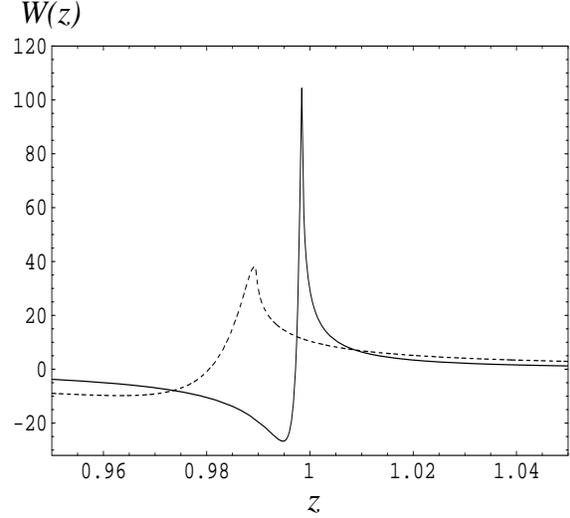,width=8cm}
\caption{Plots of $W(z)$ for $\Gamma_s=2$ (dashed) and 5 (solid). Although the peak location depends on 
$\Gamma_s$, for given $\langle\gamma'\rangle$, the function remains approximately the same
for $z>1$.}
\label{fig:wz}
\end{figure}

Since $v<1$, $W(z)$ is regular at $z=1$ and hence, the superfacial singularity at $z=1$ in (5)
can be avoided by interpolation using $W(z)\approx W(1)+(1-z)W'(1)$ with 
$W(1)=\langle\gamma^{-3}(1-v)^{-2}\rangle=\langle\gamma(1+v)^2\rangle$.
Figure \ref{fig:wz} shows plots of $W(z)$ with $\Gamma_s=2$ and 5 for $\gamma_m=2$ (corresponding
to $\langle\gamma'\rangle\approx1.2$). In the relativistic limit, one has 
$W(1)\approx4\langle\gamma\rangle \approx8\langle\gamma'\rangle\Gamma_s$. The 
peak value is $W_{\rm max}(z)\gg W(1)$.

The two relevant characteristic frequencies can be obtained from the dispersion relation:
the cut-off frequency of the Langmuir mode at 
$\omega_c=\omega_p\langle\gamma^{-3}\rangle^{1/2}$, the 
lowest allowed frequency, which can be derived by setting $z\to\infty$, and the crossing 
frequency of the LO mode $\omega_{co}\approx\omega_p\langle\gamma\rangle^{1/2}$, at 
which with decreasing plasma density, the wave evolves from the superluminal to the subluminal 
regime by crossing the light line ($z=1$). The LO mode crosses back to the superluminal regime 
above $\omega_{co}$. In the relativistic limit, one has 
$\langle\gamma^{-3}\rangle\approx1/\langle\gamma\rangle$ and then the cut-off frequency is 
$\omega_c\approx\omega_p/\langle\gamma\rangle^{1/2}$, which is lower than that for the cold 
plasma. In the mildly relativistic regime one has even a lower cut-off frequency 
$\omega_c=\omega_p\langle\gamma^{-3}\rangle^{1/2} \ll \omega_p/\langle\gamma\rangle^{1/2}$.

\section{Ray tracing in rotating magnetospheres}

Consider pulsar plasma corotating with the star and assume that the relevant length scale for the
inhomogeneities is much larger than the wavelength. It is further assumed that in the corotating 
frame the magnetic field is a static dipole. The dipole approximation may not be valid near the
light cylinder and as we consider only propagation in regions well inside the light cylinder where
deviation from the dipole approximation can be ignored.

\subsection{Hamilton's equations}

In the geometric optics approximation, the ray path is determined by a set of Hamilton's equations
for the spatial coordinate $\bf x$ and the wave vector $\bf k$. To include rotation the Hamilton's 
equations are written in the covariant form (Weinberg 1962),
\begin{eqnarray}
{dx^\mu\over ds}&=&{\partial D_M\over\partial k_\mu},\nonumber\\
{dk_\mu\over ds}&=&-{\partial D_M\over\partial x^\mu},
\label{eq:hamilton}
\end{eqnarray} 
where the metric is $g^{\mu\nu}=g_{\mu\nu}={\rm diag}(1,-1,-1,-1)$, $x^\mu=(t,{\bf x})$,
$k^\mu=(\omega,{\bf k})$, $D_M=0$ is the dispersion relation for the $M$ mode, which
is a Lorentz invariant. The solution to (\ref{eq:hamilton}) gives the ray path, represented
by $x^\mu(s)$ and $k_\mu(s)$, where $s$ is the parameterization of the ray path. The 
right-hand sides of Eq. (\ref{eq:hamilton}) can be written as
\begin{eqnarray}
{\partial D_M\over\partial k_\mu}
&=&{1\over\omega}\left(\xi_2,-\xi_1{\bf b}+2{\bf n}\right),\\
{\partial D_M\over\partial x_\mu}
&=&\left(0,\xi_1{\partial{\bf b}\over\partial {\bf x}}\cdot{\bf n}+\xi_3{\partial \ln N
\over\partial {\bf x}}\right),\\
\xi_1&=&-{\omega^2_p\over\omega^2}z^2\left[
(1-z^2)W'(z)-2zW(z)\right],\\
\xi_2&=&-{\omega^2_p\over\omega^2}\left[
2W(z)+z(z^2-1)W'(z)\right]+2n^2,\\
\xi_3&=&-{\omega^2_p\over\omega^2}(z^2-1)W(z),
\end{eqnarray}
where ${\bf b}={\bf B}/B$ is the field line direction, $n$ is the refractive index,
${\bf n}={\bf k}/k$ is the unimodule of the wave vector, and $N$ is the local plasma density.
Due to the field line curvature and density gradients, one has
$dk^\mu/ds\neq0$, which causes deviation between the wave vector $\bf k$ and the group 
velocity ${\bf v}_g=d{\bf x}/dt$.
As the ray propagates outward the coefficients, $\xi_{1-3}\propto \omega^2_p$,
decreases rapidly, at least as $1/R^3$. When the ray reaches a radial distance,
referred to as the refraction limiting radius (RLR), beyond which the refraction is no 
longer effective, $\bf k$ and ${\bf v}_g$ merges into the same direction again.

\subsection{The effect of the relativistic distribution}

The relativistic effect is described by $W(z)$ and $W'(z)$ in $\xi_1$, $\xi_2$
and $\xi_3$. In the cold plasma approximation, we have
\begin{eqnarray}
(z^2-1)W(z)&=&{z^2-1\over\Gamma^3_s(v_s-z)^2}\nonumber\\
&\approx&
{1-n^2+n^2\theta^2_{kb}\over\Gamma^3_s[(1-n)+{1\over2}n(\theta^2_{kb}+\Gamma^{-2}_s)]^2}.
\end{eqnarray}
Eq. (\ref{eq:hamilton}) reduces to the familiar ray equations in the cold plasma 
approximation (Barnard \& Arons 1986;
Petrova 2000). Refraction due to density gradients is described by
$\xi_3\propto (z^2-1)W(z)$. For $n\approx1$, one has $(z^2-1)W(z)\approx 4(\theta^2_{kb}\Gamma_s)/
[(\theta_{kb}\Gamma_s)^2+1]^2\leq 1/\Gamma_s$. In general, the effect of refraction decreases for
increasing bulk Lorentz factor, $\Gamma_s$. Refraction initially
is small at $\theta_{kb}\ll1$ and rapidly increases to reach its maximum at $\theta_{kb}\sim1/\Gamma_s$.

\begin{figure}
\psfig{file=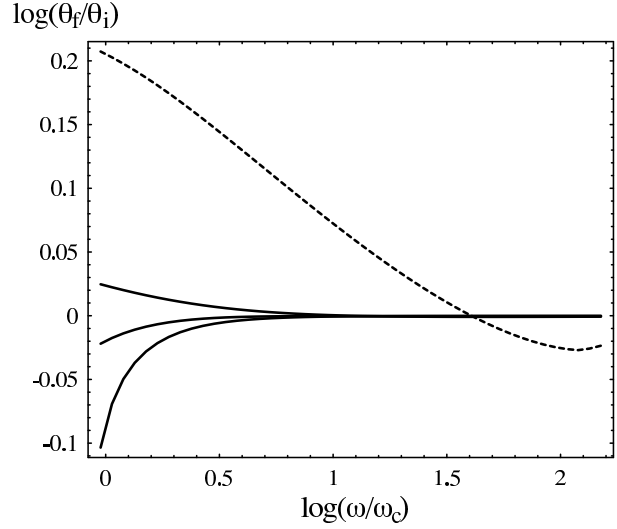,width=8cm}
\caption{Deflection of rays due to the radial gradient. The solid curves (from top to bottom)
corresponds to $\langle\gamma'\rangle=5$, 10 and 30. The refaction of rays strongly depends
on the spread of the distribution. Compared to the cold plasma case (the dashed curve),
strong refraction occurs in a much narrower frequency range towards the cut-off frequency.
The bulk Lorentz factor is $\Gamma_s=30$.}
\label{fig:defl}
\end{figure}

\begin{figure}
\psfig{file=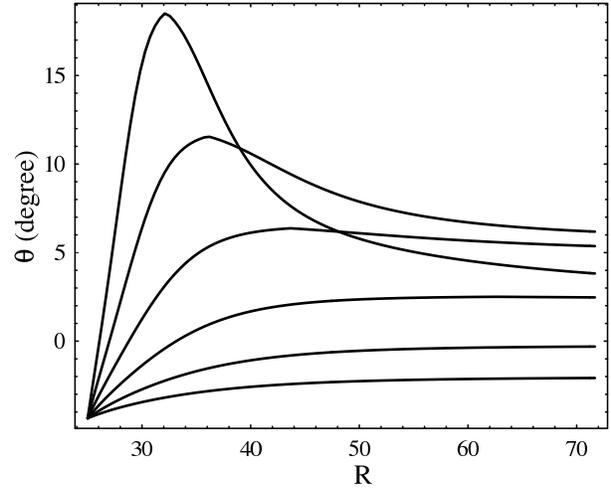,width=8cm}
\caption{Deflection of rays due to the transverse gradient. The curves
from top to bottom correspond to frequencies $\omega=10^9-10^{9.75}\,{\rm s}^{-1}$.
The radial distance, $R$, is in units of the stellar radius $R_*$.
Other parameters are $\Gamma_s=\langle\gamma'\rangle=100$, $R_0=25$, $\varepsilon_1=
\varepsilon_2=3$.}
\label{fig:trans}
\end{figure}

For an intrinsically relativistic plasma, $W(z)$ is strongly peaked near $z\approx 1$
with $W_{\rm max}(z)\sim8\langle\gamma'\rangle\Gamma_s$ (cf. Figure ~\ref{fig:wz}).
Near the peak, assuming $z\sim v_s$ one may estimate 
$(z^2-1)W(z)\sim W(1)/8\Gamma^2_s\langle\gamma'\rangle^2\approx1/\Gamma_s\langle\gamma'\rangle$,
which is smaller than that for the cold plasma by a factor of $\langle\gamma'\rangle$.
Inclusion of the relativistic distribution tends to suppress refraction in general and shifts
the strong refraction regime to $\omega<\omega_{co}$, with the strongest refraction found
near the cut-off frequency, $\omega_c$. Let $\theta$ be the colatitude of $\bf k$, which
is the propagation angle with respect to the magnetic axis. A comparison of refraction effects for 
different $\langle\gamma'\rangle$'s is shown in Figure ~\ref{fig:defl}. The ratio of the 
final (deflected) propagation angle, $\theta_f$, to the
initial propagation angle, $\theta_i$, is plotted as a function of wave frequency $\omega$ (in
$\omega_c=\omega_p/\langle\gamma\rangle^{1/2}$). The density is assumed to have a radial
gradient of $N\propto 1/R^3$ with a uniform transverse distribution. It can be seen that
the effect of refraction becomes negligible for $\omega/\omega_c>1$.

\subsection{Transverse distribution of the plasma density}

Strong refraction arises from the effect of the transverse gradient.  The major contributing term to 
Eq. (9) is from the density gradient, $\partial \ln N/\partial{\bf x}$, in the direction perpendicular to
the field line, due to the nonuniform pair cascade above the PC. In the current PC model, 
significant pair production occurs near the last open field lines
that have the smallest radius of curvature. There are few pairs produced on the field lines near 
the polar axis and no pairs on the last close field line on which the accelerating electric field
is zero. The magnetic dipole field lines in the polar coordinate in the corotating frame can be described by
$R=\sin^2\chi/\sin^2\chi_*$, where $\chi_*\leq\chi_d$ is the magnetic colatitude,  
$\chi_d=(\Omega R_*/c)^{1/2}$ is the half open angle of the PC, and $\Omega=2\pi/P$, $P$ 
is the pulse period. The radius of the open field line cone as a function of radial 
distance is $R_\perp=\chi_d R^{3/2}$, which is the characteristic length scale 
of the transverse gradient.

For a plasma streaming along the magnetic field lines, the transversely spatial distribution
of the plasma density at a given radial distance is completely determined by the  density
profile at the PC. The density profile is described by a function of the magnetic colatitude,
$\chi$, which is essentially the distribution of pair production across the PC.
The PC is assumed to be circular in the corotating frame with the opening half angle $\chi_d$.
Assuming that pair cascades have an azimuthal symmetry relative to the polar axis,   
the plasma density in the corotating frame is peaked at the field line with the colatitude
on the PC given by $\chi_c=\varepsilon_c\chi_d$ with $\varepsilon_c<1$. Specifically, the density is written as
(Barnard \& Arons 1986; Petrova 2000)
\begin{equation}
N(\chi,R)={N_*\over R^3}\left\{
\begin{array}{l}
\displaystyle{
\exp\left[-{\textstyle{1\over2}}\left({\Psi-1\over\varepsilon_1}\right)^{2p}\right]
},
\ \ \Psi<1,\\
\\
\displaystyle{
\exp\left[-{\textstyle{1\over2}}\left({\Psi-1\over\varepsilon_2}\right)^{2p}\right]
},
\ \ \Psi>1,
\end{array}
\right.
\label{eq:density}
\end{equation}
where $\Psi=\chi/\chi_cR^{1/2}\leq 1/\varepsilon_c$, $\varepsilon_{1,2}$ are 
two parameters that characterize the steepness of each side of the peak, $N_*$ is the 
density at the star's surface, the 
radial distance is in $R_*$. In polar models, the density can be written as $N_*=MN_{GJ}$, where
$N_{GJ}$ is the Goldreich-Julian density and $M$ is the pair multiplicity. Since the 
maximum pair production occurs near the last open field 
lines, one has $\chi_c\sim \chi_d$. Hence, it is reasonable to assume that $\varepsilon_2>
\varepsilon_1$. The Gaussian distribution $p=1$ in the nonrotating case was 
discussed by Barnard \& Arons (1986), also Petrova (2000). We consider a more general case
including higher density gradients $p>1$. 

Figure~\ref{fig:trans} shows ray paths in the $\theta$-$R$ plane. The rays are emitted at 
$R_0=25$ with the colatitude $\theta=-5^\circ$, where the negative angle corresponds to 
the trailing side of the magnetic axis, and they are refracted by the transverse gradient. 
Notice that refraction is sensitive to frequency: the lower frequency the stronger refraction. 
Because of strong deflection the first three rays (from top) cross the magnetic axis into the leading side. 

\subsection{Rotating medium}

Rotation can be included in ray tracing as follows.  Assuming that the wave is
emitted in the field line direction in the corotating frame in which the magnetic
field is a dipole by assumption. The ray path is calculated as a function of time $t$ in the 
observer's frame. Specifically, this involves solving the Hamiltonian equations with a 
sequence of transformations, first rotation and then the Lorentz transform,
\begin{eqnarray}
T&=&T_2(\alpha)T_3(\Omega t)T_2(\alpha)^{-1},\label{eq:rotation}\\
L&=&L({\bf V}_r),
\label{eq:lorentz}
\end{eqnarray}
where the 3-axis is along the pulsar spin axis (${\bf \Omega}$),
the magnetic pole is assumed to be in the 1-3 plane at $t=0$, 
$T_2$ and $T_3$ are the rotation with respect to the 2- and 3-axis, respectively,
$\alpha$ is the inclination angle (relative to the rotation axis), 
${\bf V}_r={\bf \Omega}\times{\bf R}$ is the co-rotation 
velocity at the radial distance, $R$, $\bf \Omega$ is the angular velocity of the rotation,
and $L({\bf V}_r)$ is the Lorentz transform. Eq. (\ref{eq:rotation}) represents a combined 
rotation, i.e. an anticlockweise rotation about the 2-axis by $\alpha$ so that the magnetic 
pole is aligned with ${\bf \Omega}$, then rotation about the 3-axis by $\Omega t$, followed by a 
clockweise rotation about the 2-axis by $\alpha$. In the observer's frame we have
\begin{eqnarray}
{d\tilde{x}^\mu\over ds}&=&
T^{\mu}_\nu L^\nu_\lambda{\partial D_M\over\partial k^\lambda},
\label{eq:rayEq2a}\\
{d\tilde{k}^\mu\over ds}&=&-
T^{\mu}_\nu L^\nu_\lambda{\partial D_M\over\partial x^\lambda}.
\label{eq:rayEq2b}
\end{eqnarray}
Ray trajectories can be obtained by numerically integration of Eq. (\ref{eq:rayEq2a}) 
and (\ref{eq:rayEq2b}) through the refraction zone,  the region between
the emission radius to the RLR. At the RLR, the propagation time, $t$, and the 
final wave vector, $\tilde{\bf k}_f$, are determined. A rotational transform, determined
by $t-t_0$, where $t_0$ is the time that a vacuum wave would take to propagate through 
the region, is then applied to $\tilde{\bf k}_f$. One can calculate the final wave 
vectors for all rays and the distribution of a bunch of rays can be determined at a 
particular instance.

\section{Application to pulsars}

The specific mechanism for radio emission is not well understood.
One possibility is that the radiation is produced through linear wave-particle 
interactions. This is only possible for subluminal waves (with phase velocity less
than $c$). The LO mode can be subluminal only at frequencies near the cross frequency
$\omega_{co}$ within a small range of propagation angle, $\theta_{kb}\leq\theta_m\approx
(2\langle\gamma'\rangle)^{1/2}\omega_p/\Omega_e\ll1$ (Melrose \& Gedalin 1999;
KML). The radio emission can be produced in other types of wave and then converted to
propagate modes. In this case it is possible that the radio emission is in the superluminal
modes with frequency below the cross frequency $\omega\leq\omega_{co}$. If this is the case, refraction
may have a significant effect on wave propagation.

\begin{figure}
\psfig{file=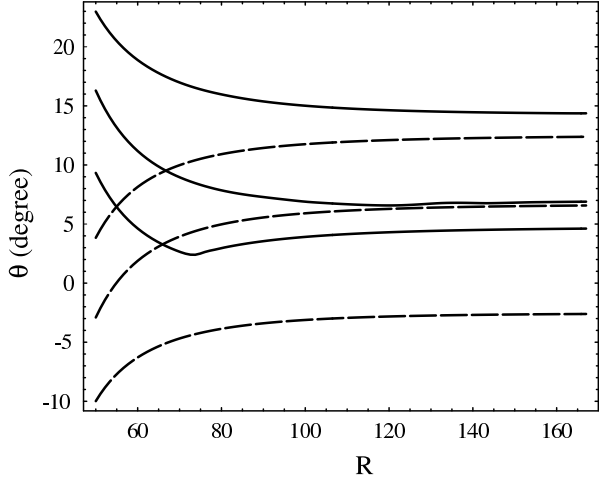,width=8cm}
\caption{Asymmetry due to differential refraction caused by star's rotation.
The solid and dashed curves represent respectively the ray paths that orginate
from the leading and trailing sides of a dipole with the inclination $\alpha=\pi/2$.
The rotation is in the direction of positive $\theta$. The parameters are as in Figure 3 but  
with the emission radius $R_0=50$; $p=2$; $(P,\dot{P})=(0.1\,{\rm s},10^{-13})$;
$\omega=10^{9.2}\,{\rm s}^{-1}$; the pair multiplicity is assumed to be $M=10^4$.}
\label{fig:asym1}
\end{figure}

\begin{figure}
\psfig{file=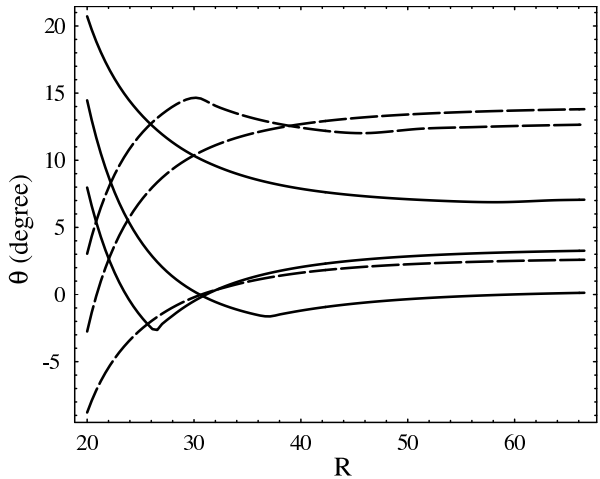,width=8cm}
\caption{Differential refraction as in Figure \ref{fig:asym1} with strong deflection.
The asymmetry is easiest to be seen in strongly deflected rays. The emission
radius is $R_0=20$.}
\label{fig:asym2}
\end{figure}

\subsection{Emission beam}

Since the intensity distribution in the emission region is not known, one
considers only the case of emission from a particular height with
a 2-dimensional distribution in intensity (e.g. Petrova 2000; FLM).
The initial intensity profile in the co-rotating frame is assumed to
follow the density profile and be Gaussian in colatitudes with axisymmetry relative to the pole,
that is,
\begin{equation}
I_0(\theta_0,R_0)=\exp\left[
-{\textstyle{1\over2}}\left({\Psi_0-1\over\varepsilon}\right)^2\right],
\label{eq:initial_intensity}
\end{equation}
where $\varepsilon=(\varepsilon_1,\varepsilon_2)$ are defined in (\ref{eq:density}),
$\Psi_0=\chi_0/\chi_cR^{1/2}_0\leq 1/\varepsilon_c$, $\chi_0$ is the colatitude of the 
emission point. We have $\chi_0\approx 2\theta_0/3$, where $\theta_0$ is the corresponding 
colatitude of $\bf k$ at the emission origin. Without refraction the intensity distribution 
(\ref{eq:initial_intensity}) would simply lead to a single emission cone.  Because 
of refraction the distribution of 
rays changes as they propagate away from the emission region up to the RLR beyond which the 
refraction is no longer important. The emergent beam is obtained by calculation of the ray 
distribution at the RLR.

\subsection{Asymmetry in pulse profiles}

Many pulsars have multicomponent pulse profiles, commonly characterized by 
conal structures (e.g. Lyne \& Manchester 1988; Rankin 1993)
The conal components appear to shift toward the rotation direction,
which is interpreted as the effect of aberration or retardation (e.g. Blaskiewicz, Cordes \& 
Wasserman 1991; Gangadhara \& Gupta 2001; Gupta \& Gangadhara 2003). A special
emission geometry such as nested emission cones is often evoked to explain the conal 
structure of the profile. Although pair production above the PC is expected to be rather
nonuniform it is not clear how such a regular nested conal structure forms. In observations, 
conal components generally show asymmetry in intensities, i.e. some pulsars have a stronger leading 
component, while some others have a stronger trailing component. Our model suggests that differential 
refraction can produce such asymmetric pulse profiles without appealing to the nested conal structure 
at the emission origin. 

\begin{figure}
\begin{center}
\psfig{file=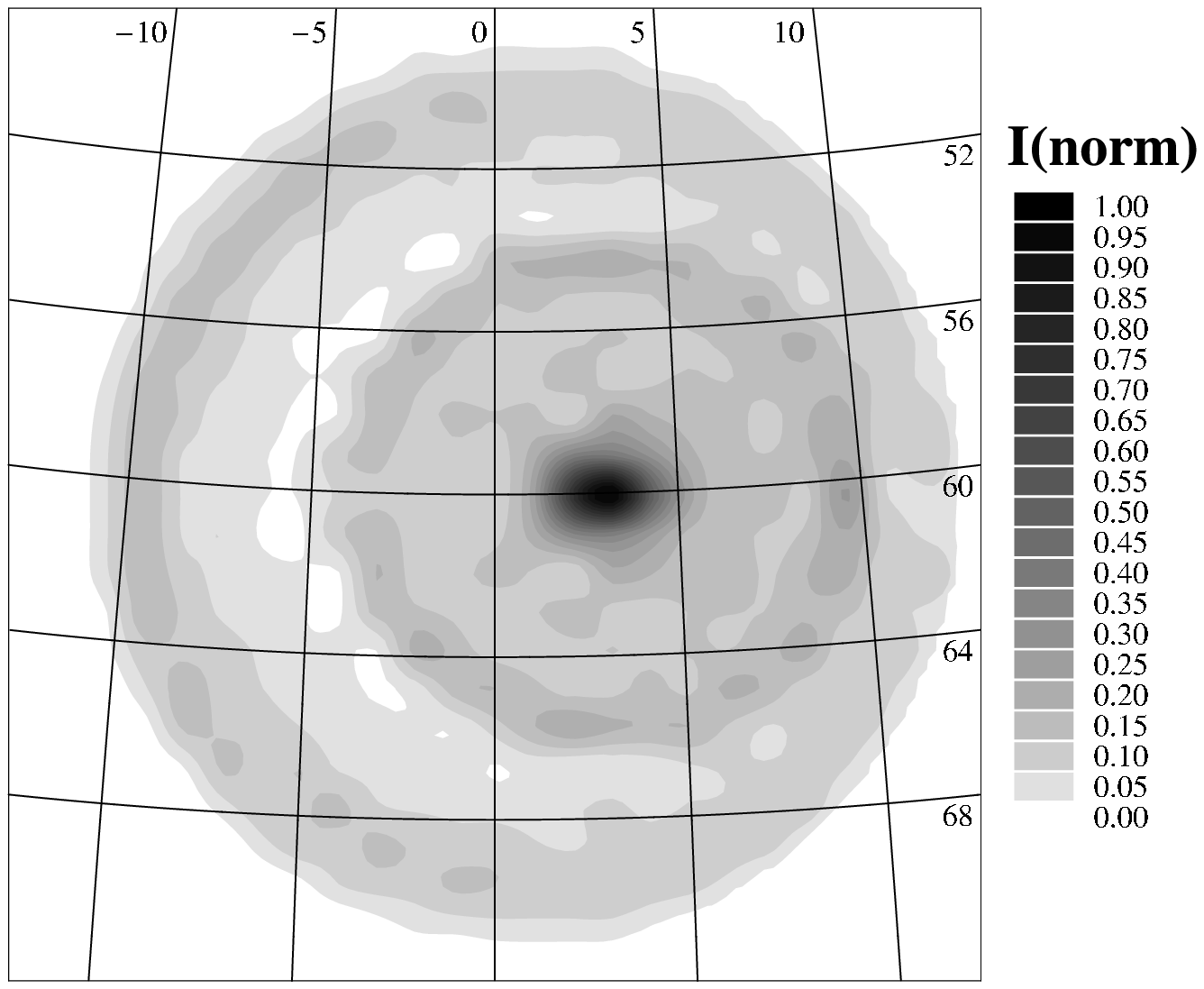,width=8cm}
\psfig{file=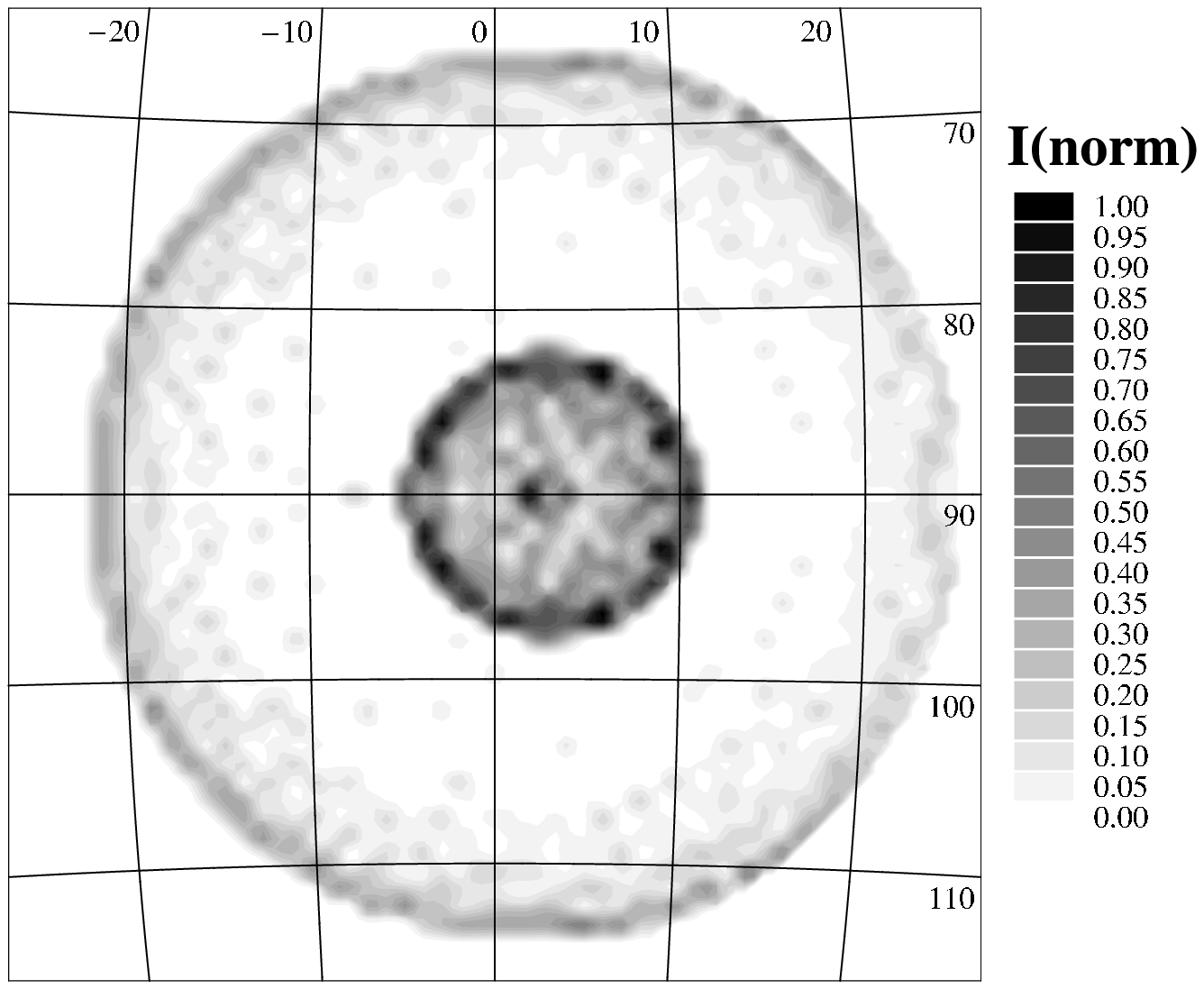,width=8cm}
\end{center}
\caption{Beam intensity profiles. The latitude and longitude (in degree) correspond respectively
to the viewing angle (relative to the rotation axis) and the pulse phase.
The upper: $p=2$, $\Gamma_s\approx\langle\gamma\rangle\approx16$,
$\alpha=60^\circ$, $P=1\, \rm s$. The lower: $p=1$, $\Gamma_s\approx\langle\gamma\rangle\approx100$,
$\alpha=90^\circ$, $P=0.1\, \rm s$. The zero pulse phase corresponds to that the line of 
sight is in the plane of the magnetic axis and rotation axis. The positive longitude corresponds 
to the direction of rotation.
}
\label{fig:beam}
\end{figure}

Figures \ref{fig:asym1} and \ref{fig:asym2} show asymmetry in refracted ray paths in the observer's
frame. One assumes the density profile with $p=2$ and 
$(\varepsilon_1,\varepsilon_2,\varepsilon_c)=(0.64,0.6,0.8)$.
The three pairs (solid and dashed) of ray path originate inside the critical field lines
($\chi_*=\chi_c$) and are initially symmetric on emission in the
corotating frame, with the intensity distribution given by (19). 
In the co-rotating frame, rays emitted at the same colatitude on the 
opposite sides of the pole are subject to different plasma gradients,
and due to the differential refraction their paths are no longer symmetrical. The rays originating on the 
leading side of the pole are focused back towards the center earlier on their paths than the ones
on the trailing side. If one decreases the inclination angle $\alpha$, the rotation effect is reduced. 
The ray paths in Figures \ref{fig:asym1} and \ref{fig:asym2} appear to focus since they originate
inside the critical field lines. If the density distribution is strongly peaked near the last open
field lines and all radiation is produced inside the critical field lines, refraction can lead 
to a narrower pulse width.  

\begin{figure}
\psfig{file=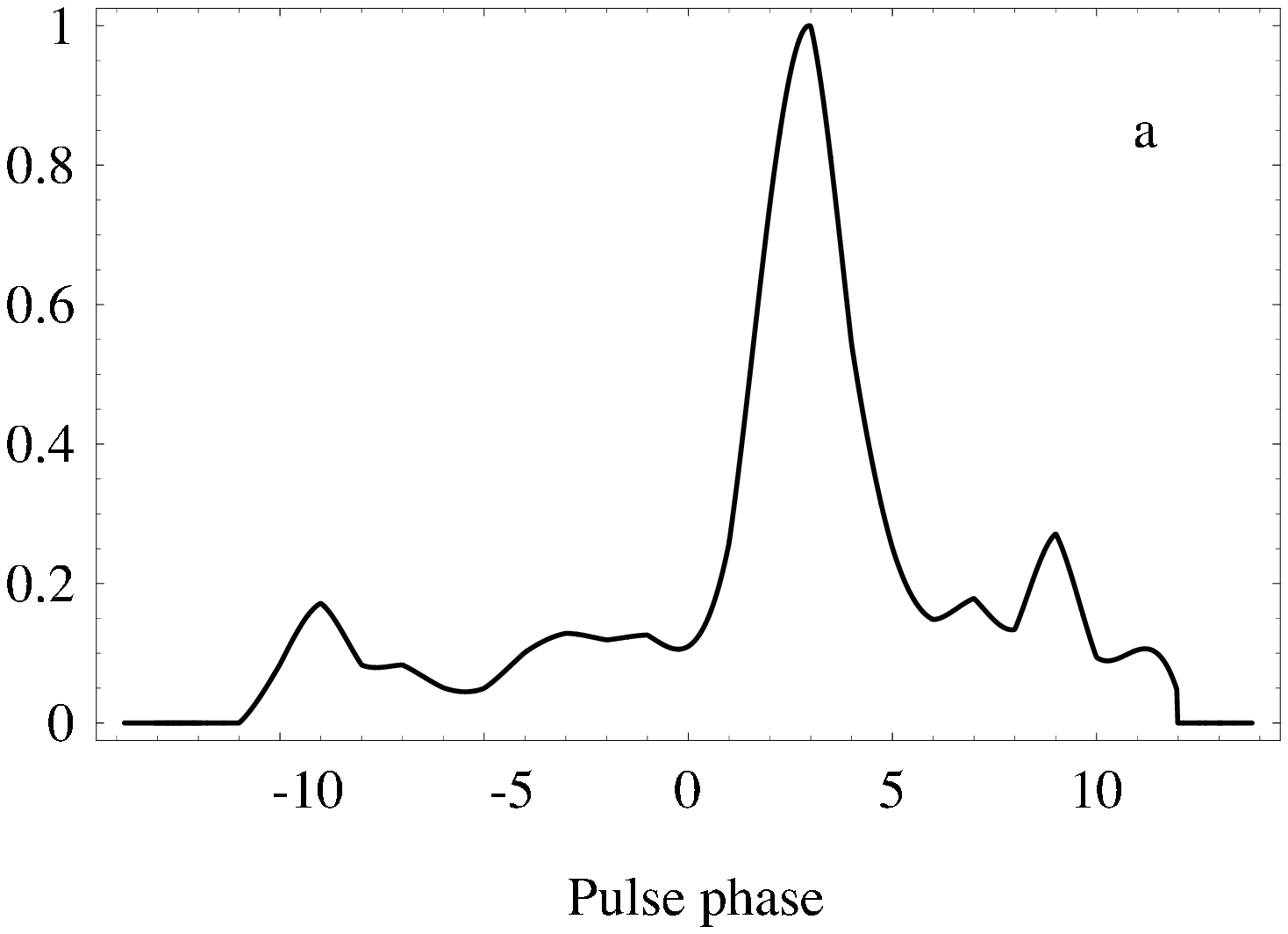,width=8cm}
\psfig{file=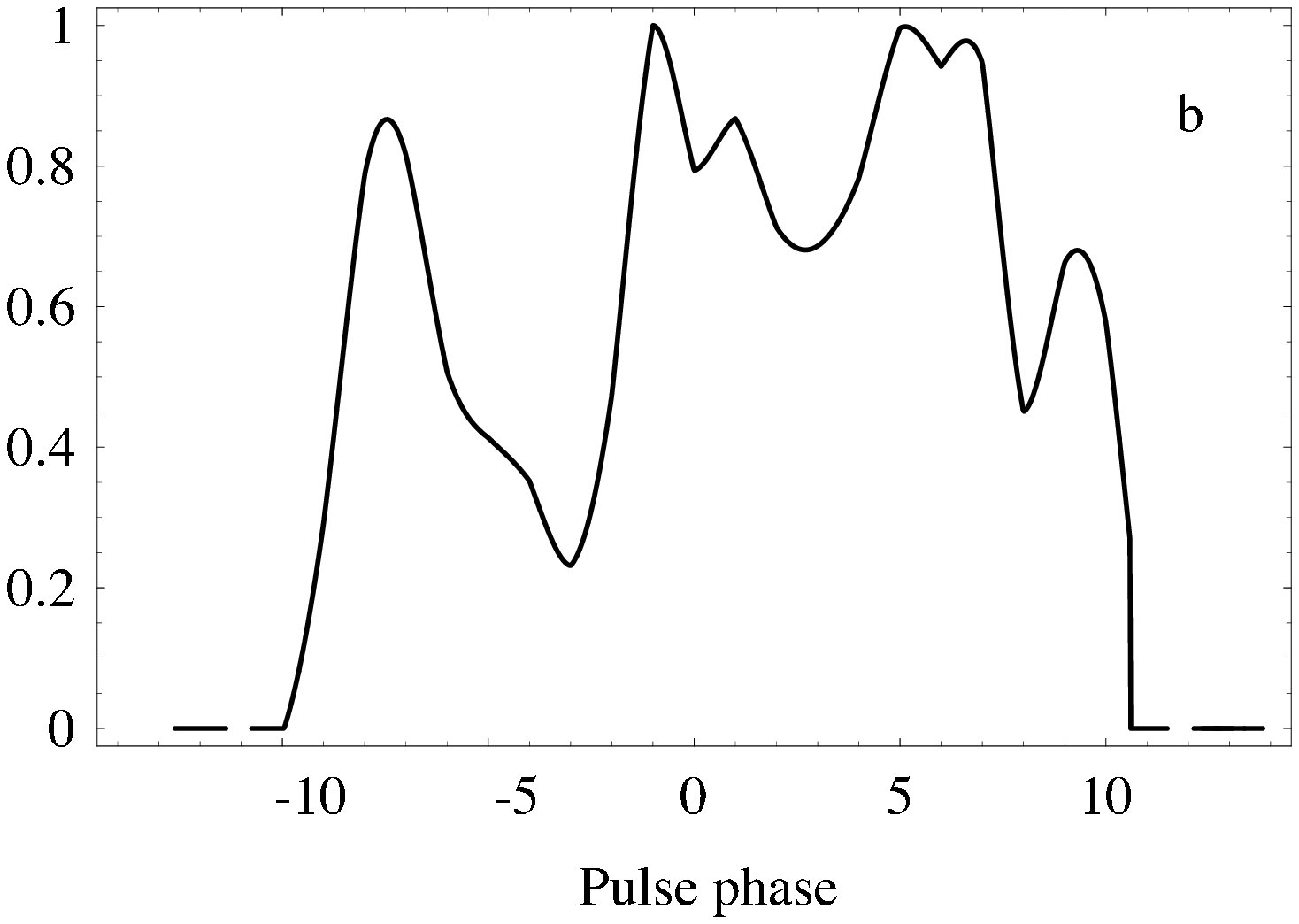,width=8cm}
\psfig{file=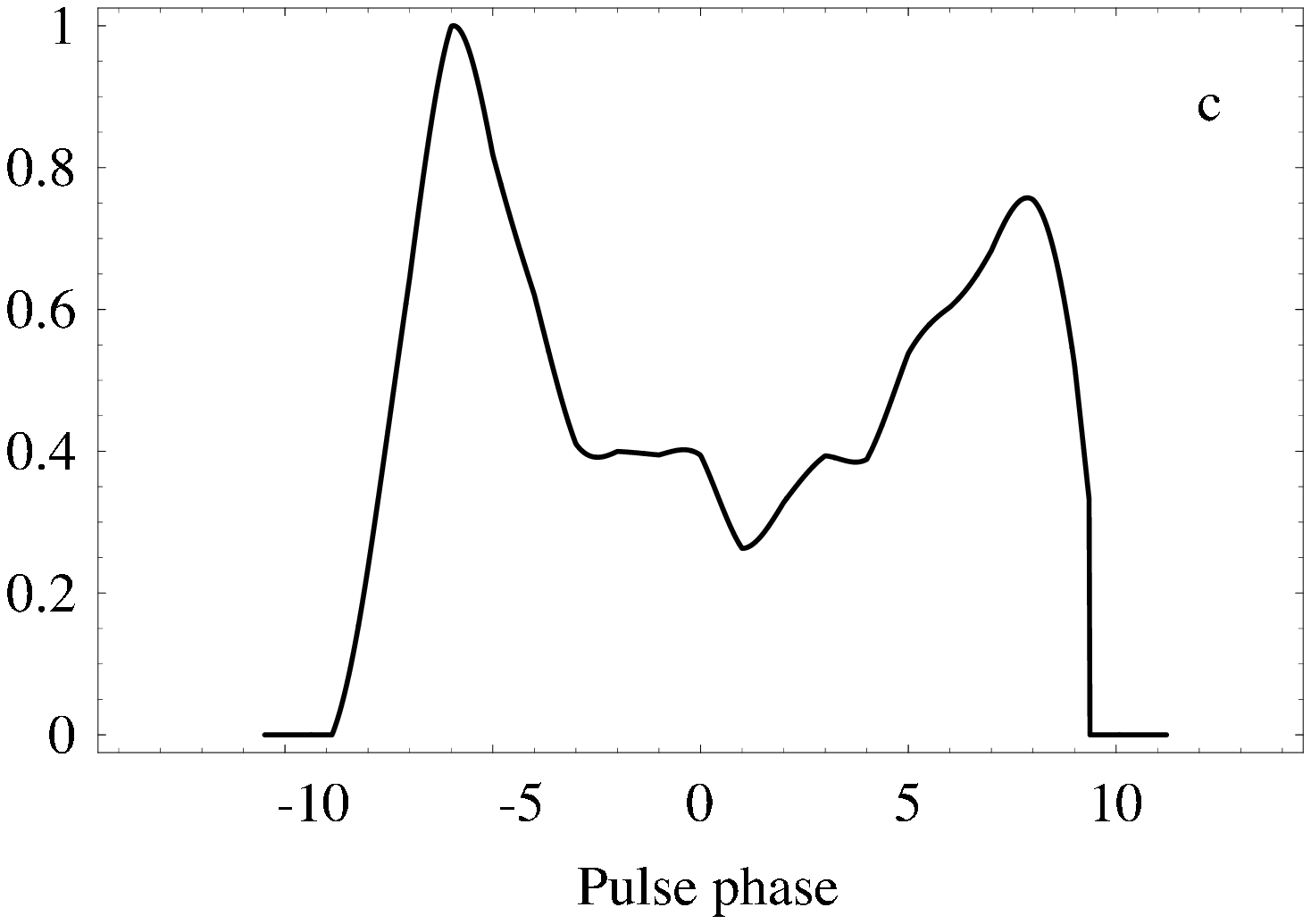,width=8cm}
\caption{Pulse profiles for three different viewing angles for the emission beam
in the upper panel of Figure~\ref{fig:beam}.  a: $i=60^\circ$; b: $55^\circ$; 
and c: $53^\circ$. The intensities in relative scale (with the maximum intensity in 
each case being normalized to 1) are plotted against the pulse phase (degree). 
}
\label{fig:profile}
\end{figure}

The beam intensity profiles in the observer's frame
are shown in Figure \ref{fig:beam} for the plasma density profiles 
of $p=1$ and $p=2$. The intensity profiles are obtained by evaluating the ray distribution at the RLR.
The longitude (in degree) is the pulse phase, with the zero pulse phase corresponding to
the line of sight in the plane of the magnetic axis and rotation axis. The positive longitude corresponds
to the leading side, which is opposite to the definition adoped in Blaskiewicz et al (1991). In their Figure 3,
the negative phase corresponds to the leading direction and the zero pulse phase is where
the position angle variation is the steepest. We assume $\omega=10^9\,{\rm s}^{-1}$,
$M=10^4$, $\dot{P}=10^{-15}$, $R_0=20$ and $(\varepsilon_1,\varepsilon_2,\varepsilon_c)=(0.4,0.35,0.8)$ for
$p=1$ and $(\varepsilon_1,\varepsilon_2,\varepsilon_c)=(0.64,0.6,0.8)$ for $p=2$.
The conal beam is split into two nested cones due to the refraction effect, with emission
produced inside $\chi_c$ bending towards the magnetic axis forming the inner cone and emission 
produced outside $\chi_c$ deflected away forming the outer cone. Apart from two main nested cones,
some more complicated conal structures at low-level intensities can also be seen.
When the refraction is strong, as in the case of steep density gradient ($p=2$), the rays 
forming the inner cone cross the magnetic axis producing bifurcation of the core.  Examples 
of multiple component profiles resulted from 
the bifurcation are shown in Figure~\ref{fig:profile}. The profiles are obtained from the emergent 
emission beam in the upper subfigure in Figure \ref{fig:beam} for three different viewing angles 
(the angle between the line of sight and the rotation axis).
The first profile, which is obtained for the viewing angle $i=\alpha=60^\circ$,
has a dominant core structure as the result of rays being focused toward the 
magnetic pole axis. All three profiles have nested conal structures. Since rays originating 
on the leading and trialing side are subject to different refraction, the distribution of 
conal components skews toward the rotation direction with uneven intensities for the leading and 
trailing components. In the absence of refraction, the aberration would advance 
the emission toward the leading direction (positive longitude) by a phase of only about
$R_0/R_{LC}\approx 2.4$ degree for $R_0=20$. The differential refraction appears to shift the 
emission further in the leading direction (cf.Figure~\ref{fig:beam}). In our ray tracing model, 
the ray path is not only affected by the aberration but also by differential refraction. 
The latter effect is dominant for the emission from low altitudes. 

\subsection{Frequency dependence}

As in the cold plasma model, refraction strongly depends on frequency and is significant 
at lower frequencies relative to the plasma frequency. The observational implication depends on
emission models in which the wave frequency can be related to the plasma parameters. 
If one assumes that the radiation frequency is related to the relativistic plasma frequency, 
say $\omega=\eta\omega_c$, where $\eta>1$ is a model-dependent parameter. The 
frequency dependence is then given by a function of $\omega_c$,
which is mainly determined by the model for the density gradient. For the transverse 
gradient model discussed in Sec. 3.3, one has stronger refraction at higher frequencies 
(corresponding to a higher $\eta\omega_c$). This is because the emission region with 
high $\omega_c$ must have  a narrower density profile with a larger gradient and 
the ray is then subject to a stronger refraction (e.g. Petrova 2000).

Although the hypothesis of emission at the frequency fixed relative to the plasma
frequency predicts a radius-to-frequency relation, $\omega\sim R^{-3/2}$, roughly  
consistent with observations (Petrova 2000), there is no widely accepted model for the 
emission mechanism that predicts such a frequency relation. This assumption requires that 
the radiation to be produced at a variable range of radial distances, which would complicate 
the frequency to emission radius relation (Weltevrede et al. 2003).  

\subsection{Cyclotron absorption}

A ray that propagates outward eventually reaches the cyclotron resonance
region and cyclotron absorption occurs when the wave frequency equals the cyclotron frequency in the 
electron rest frame. In general the resonance region is located well above the RLR. 
Whether the resonance zone is located within the light cylinder depends on the bulk Lorentz factor
of the plasma as well as the relativistic spread. It was shown by several authors that
the cyclotron resonance condition can be satisfied inside the light cylinder for some pulsars
(Blandford \& Scharlemann 1976; Mikhailovskii et al. 1982; Luo \& Melrose 2001; FLM). 
If cyclotron absorption occurs it can distort the pulse profile through differential
absorption (Luo \& Melrose 2001; FLM).  Since the optical depth for the absorption
is $\tau\propto\theta^2N$, rays that originate on the leading and trailing sides propagate
along different paths with asymmetry in $\theta$ and the density, leading to
differential absorption. 

\section{Conclusions}

The effect of refraction on wave propagation in rotating magnetospheres is considered
including both effects of relativistic distribution and rotation. Since the
relativistic plasma dispersion function is peaked at around $z\approx1$ and remains 
small otherwise, inclusion of the intrinsic relativistic effect tends to suppress of refraction
at frequencies $\omega\geq\omega_c$. Refraction is sensitive to the plasma 
density profile in the transverse direction and causes rays to focus and bifurcate,
which is qualitatively similar to the result from the nonrotation case (e.g. Petrova 2000). 
The focusing effect tends to produce a core component even when the emission region has a conal
geometry. The bifurcation leads to split of the emission cone into two or more nested cones, 
giving rise to profiles with multiple components.  

One of the distinct features of the model discussed here is the prediction of asymmetry in the 
pulse profile. The conal components are skewed toward the rotation direction, which is qualitatively in 
agreement with recent observations (e.g. Gangadhara \& Gupta 2001). The predicted profiles
also show asymmetry in relative intensities between the leading and trailing components, which 
are common in observed pulse profiles. It is suggested that similar asymmetry seen in observations
can be produced by combination of aberration and the differential refraction due to rotation. The latter
is the dominant cause for the asymmetry if the radiation is produced at low altitudes.

Two other modes that are not discussed here include the X mode and low-frequency Alfv\'en 
mode. In general, pulsar plasma is gyrotropic due to that the electron and positron distributions
are not identical. Although the X mode is not purely transverse, with a small longitudinal component 
of along the magnetic field, one expects refraction of the X mode to be weaker for the LO mode. 
Inclusion of the X mode would lead to pulse profiles of the two modes that are displaced
with one mode dominating any particular part of the pulse longitude. The observational implication
of such a displaced profile needs to be explored. The low-frequency Alfv\'en mode can be refracted in the 
low-frequency regime. However, the wave becomes subluminal as it propagates to underdense regions  
and is subject to strong damping.

Pair production above the PC can be oscillatory over the characteristic time scale about
$\Delta t_0=h_0/c$, where $h_0$ is the typical length of polar gap acceleration. 
The pulsar plasma formed from the cascade may consist of many outflowing clouds
(e.g. Usov 1987; Asseo \& Melikidze 1998). For $h_0\sim R_*$, we have 
$\Delta t_0\sim 10^{-4}\,\rm s$. Since this time can be shorter than the ray propagation 
time, further work on the wave propagation including the time-dependence of the medium 
is needed.

\section*{Acknowledgment}

We thank Don Melrose and Simon Johnston for useful discussion.

\bibliography{pnb}
\bibliographystyle{mn}
\label{lastpage}

\end{document}